\documentclass[citeautoscript, superscriptaddress,twocolumn,showpacs,prb,aps,10pt]{revtex4}

\usepackage{graphicx} \usepackage{amsmath, amssymb, bm}
\usepackage{dcolumn} \usepackage{color}
\usepackage{eso-pic}
\usepackage{fix-cm}

\usepackage{multirow}
\usepackage{sidecap}

\usepackage{marginnote}
\usepackage{ulem} 

\newcommand{\sss}{\scriptscriptstyle}
\newcommand{\sst}{\scriptstyle}

\newcommand{\stext}[1]{\sss \text{#1} \sst}

\renewcommand{\emph}[1]{\textit{#1}}

\begin{document}
\title{The Lyddane-Sachs-Teller relationship for polar vibrations in materials with monoclinic and triclinic crystal systems}

\author{Mathias~Schubert}
\email{schubert@engr.unl.edu}
\homepage{http://ellipsometry.unl.edu}
\affiliation{Department of Electrical and Computer Engineering and Center for Nanohybrid Functional Materials, University of Nebraska-Lincoln, U.S.A.}
\affiliation{Leibniz Institute for Polymer Research, Dresden, Germany}
\date{}

\begin{abstract}
A generalization of the Lyddane-Sachs-Teller relation is presented for polar vibrations in materials with monoclinic and triclinic crystal systems. The generalization is derived from an eigen dielectric displacement vector summation approach, which is equivalent to the microscopic Born-Huang description of polar lattice vibrations. An expression for a general oscillator strength is also described for materials with monoclinic and triclinic crystal systems. A generalized factorized form of the dielectric response characteristic for monoclinic and triclinic materials is proposed. The generalized Lyddane-Sachs-Teller relation is found valid for monoclinic $\beta$-Ga$_2$O$_3$, where accurate experimental data became available recently from a comprehensive generalized ellipsometry investigation. Data for triclinic crystal systems can be measured by generalized ellipsometry as well, and are anticipated to become available soon and results can be compared with the generalized relations presented here. 
\end{abstract}
\pacs{63.20.-e;78.30.Am;78.20.Ci;78.30.-j} \maketitle

The Lyddane-Sachs-Teller (LST) relation~\cite{Lyddane41} sets two important ratios equal for a material with polar vibrations. The square of the ratio of the frequency of longitudinal optic lattice vibrations (phonons) ($\omega_{\text{{LO}}}$) to the frequency of transverse optical lattice vibration ($\omega_{\text{{TO}}})$ for long wavelengths and at negligible wave vector equals the ratio of the dielectric permittivity at zero frequency $\varepsilon_{\stext{DC}}$ with the dielectric permittivity at frequencies above the TO and LO vibrations where the material is widely transparent ($\varepsilon_{\infty}$)\footnote{The region of transparency should be a region with little or no dispersion, for example, as exhibited by a wide band gap material in the visible spectral range. The LST relation then can be used to connect the index of refraction in the visible or near infrared spectral region with its counter part at zero frequencies through all LO and TO frequencies.} 

\begin{equation}\label{eq:LSTsimple}
\frac{\varepsilon_{\stext{DC}}}{\varepsilon_{\infty}}=\left(\frac{\omega_{\stext{LO},l}}{\omega_{\stext{TO},l}}\right)^2.
\end{equation}  

The LST relation is a fundamental statement, and can be found in many text books on condensed matter physics and semiconductor optics~\cite{Pidgeon80,Wolfe89,Kittel2009,Klingshirn95,Yu99,SchubertIRSEBook_2004,GrundmannBook,Fujiwara_2007}. The LST relation has been used extensively, either to predict a missing parameter out of the set of fundamental four, $\varepsilon_{\stext{DC}}, \varepsilon_{\infty}, \omega_{\stext{LO}}, \omega_{\stext{TO}}$, or to check for consistency among experimentally and/or computationally obtained phonon mode and dielectric permittivity parameters. The LST relation has been expanded previously to include situations where multiple branches of phonon modes occur, and the role of poles and zeros in the complex plane to describe the dielectric response functions was identified~\cite{Berreman68,Gervais74}

\begin{equation}\label{eq:LSTproduct}
\frac{\varepsilon_{\stext{DC}}}{\varepsilon_{\infty}}=\prod^{N}_{l=1}\left(\frac{\omega_{\stext{LO},(l)}}{\omega_{\stext{TO},(l)}}\right)^2,
\end{equation} 

\noindent where $N$ denotes all long wavelength active polar vibration modes with polarization parallel to a certain but fixed crystal direction~\cite{Gervais74,SchubertPRB61_2000,SchoecheJAPTiO22013}. Derived for isotropic materials and most commonly applied to isotropic materials, the relation has been found correct for anisotropic materials whose major axes of polarization align with orthogonal axes~\cite{SchubertPRB61_2000,SchoecheJAPTiO22013,SchubertIRSEBook_2004}. Such situations include materials with cubic, hexagonal, trigonal, tetragonal, and orthorhombic crystal systems~\cite{Kleber}. The LST relation, however, is not valid anymore for materials with monoclinic or triclinic crystal systems. Contemporary semiconductor materials are cubic (for example, diamond structure silicon, and zincblende-structure group-III phosphides, arsenides and selenides), or hexagonal (for example wurtzite-structure group-III nitrides). Very recently, the monoclinic phase of metal-oxide $\beta$-Ga$_2$O$_3$ (gallia) has emerged as potential candidate for use in high-power transistors and switches due to a very large electric break down field value of 8 MVcm$^{-1}$\cite{KoheiJCG2013}. First devices exhibited excellent characteristics such as a nearly ideal pinch-off of the drain current, an off-state breakdown voltage over 250 V, a high on/off drain current ratio of around 10$^{4}$, and small gate leakage current~\cite{Higashiwakipssa2014}. A generalized ellipsometry analysis~\cite{Schubert96,Schubert03a} of phonon modes and free charge carrier properties in $\beta$-Ga$_2$O$_3$ was reported very recently. Traditional approaches valid for cubic, hexagonal, trigonal, tetragonal, and orthorhombic crystal systems, for derivation of phonon modes and dielectric constants from the long wavelength dielectric function  behavior of materials, require modification to account for the monoclinic character of $\beta$-Ga$_2$O$_3$~\cite{SchubertPRB2016}. Few reports exist on long wavelength characterization of monoclinic CdWO$_4$~\cite{JellisonPRB2011CdWO4}, CuO~\cite{KuzmenkoPRB2001CuOIR}, and MnWO$_4$~\cite{MoellerPRB2014MnWO4} where the application of the LST relation was not discussed. Virtually no information is available on triclinic materials, and which appears as a widely uncharted field of condensed matter physics.

In this paper, a generalization of the LST relation is derived, which is valid for materials of all crystal systems, including monoclinic and triclinic. The paper follows a derivation of a general expression of the dielectric function tensor for materials with polar vibrations. A simple superposition of eigen dielectric displacement polarizability functions and their vector character leads to complex tensor description, from where a general LST relation is obtained. The derivation is equivalent to the microscopic description given by Born and Huang~\cite{Born54}, however, the derivation is straight forward and produces simple expressions for the dielectric tensor components which can be compared conveniently with experimentally accessible polarized reflectance and transmittance data.

\begin{figure}[!tbp]
  \begin{center}
    \includegraphics[width=0.3\linewidth]{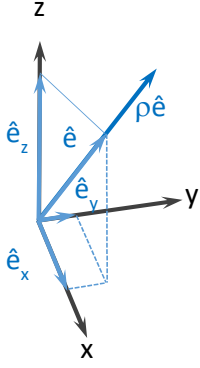}
    \caption{Unit vector $\hat{e}$ characteristic for a dielectric eigen polarizability vibration $\mathbf{P}_{\hat{e}}$ whose frequency response is rendered by a complex-valued response function $\varrho$.}
    \label{fig:eigendisplacement}
  \end{center}
\end{figure}

Vibration modes which can be excited by long wavelength electromagentic waves (long wavelength active phonon modes) in materials can be represented as intrinsic dielectric polarizations (eigen dielectric displacement modes). Each mode produces an electric dipole charge oscillation. The dipole axis can be associated with a characteristic vector (unit eigen displacement vector $\hat{e}$). Optical properties of all crystal systems can be described: monoclinic, triclinic, orthorhombic, tetragonal, hexagonal, trigonal, and cubic. The mutual orientations of the $N$ eigenvectors, and the frequency responses of their eigen displacements determine the optical character of a given, dielectrically polarizable material. For certain or all frequency regions, the optical behavior may be isotropic, uniaxial, or biaxial~\cite{Born2002}. Within the frequency domain, and within a Cartesian system with unit axes \textbf{x}, \textbf{y}, \textbf{z}, the dielectric polarizability $\mathbf{P}$ under the influence of an electric phasor field $\mathbf{E}$ along $\mathbf{\hat{e}}=\hat{e}_x\mathbf{x}+\hat{e}_y\mathbf{y}+\hat{e}_z\mathbf{z}$ is then given by a complex-valued response function $\varrho$ (Fig.~\ref{fig:eigendisplacement})~\cite{SchubertPRB2016}

\begin{equation}
\mathbf{P}_{\hat{e}}=\varrho(\mathbf{\hat{e}}\mathbf{E})\mathbf{\hat{e}}.
\end{equation}

\noindent Function $\varrho$ must satisfy causality and energy conservation requirements, i.e., the Kramers-Kronig (KK) integral relations and $Im \{ \varrho \} \ge 0, \forall$ $\omega \ge 0$~\cite{Dressel_2002,Jackson75}. The eigenvectors are located along certain, fixed spatial directions within a given sample of material. Explicit coupling between different eigen displacement modes, which may lead to description of chiral properties, are ignored here. The linear polarization response of a material with $N$ eigen displacement modes is then obtained from summation

\begin{equation}\label{eq:sumP}
\mathbf{P}=\chi\mathbf{E}=\sum^{N}_{l=1}\mathbf{P}_{\hat{e_l}}=\sum^{N}_{l=1}\varrho_{(l)}(\mathbf{\hat{e}}_l\otimes\mathbf{\hat{e}}_l)\mathbf{E},
\end{equation}

\noindent where $\otimes$ is the dyadic product. The energy (frequency) dependent contribution to the long wavelength polarization response of an uncoupled electric dipole charge oscillation is commonly described using a Kramers-Kronig consistent Lorentzian broadened oscillator function~\cite{SchubertIRSEBook_2004,HumlicekHOE}

\begin{equation}\label{eq:rho}
\varrho_{(l)} \left(\omega\right)=\frac{A_{(l)}}{\omega^2_{\stext{TO},(l)}-\omega^2-i\omega\gamma_{(l)}},
\end{equation}

\noindent where $A_{(l)}$, $\omega_{\stext{TO},(l)}$, and $\gamma_{(l)}$ denote the amplitude, resonance frequency, and broadening parameter of a vibration mode with transverse optical (TO) character, $\omega$ is the frequency of the driving electromagnetic field, and $i^2=-1$ is the imaginary unit. The index $l$ numerates the contributions of all independent dipole oscillations. It is required here that $\omega_{\stext{TO},(l)} > 0\; \forall l$.\footnote{The Drude quasi-free electron model~\cite{Drude04} is equivalent to Eq.~(\ref{eq:rho}) with $\omega_{\stext{TO},(l)}=0$ because no restoring force exists on the free charge carriers in this model. Hence, analogous to the derivation of the LST relation in its original paper, for the generalization discussed here free carrier contributions must be excluded.} The field phasors displacement $\mathbf{D}$, and $\mathbf{E}$ are related by the dielectric function tensor ($\varepsilon_0$ is the vacuum permittivity)

\begin{equation}\label{eq:eps}
\mathbf{D}=\varepsilon_0\left(\varepsilon_{\infty}+\chi\mathbf{E}\right)=\varepsilon_0\varepsilon\mathbf{E},
\end{equation}

\noindent where a symmetric tensor $\varepsilon_{\infty}$ may account for the high frequency appearance of $\varepsilon$. The high frequency limit here is meant as a frequency region with frequencies sufficiently large against the vibration modes summed over in Eq.~(\ref{eq:sumP}), and yet small against potential other electronic polarizabilities whose transition energies are at even higher frequencies. One may inspect $\det\{\varepsilon(\omega)\}$ for $\omega \rightarrow 0$ and for $\omega \rightarrow \infty$. Six real-valued physical material parameters may be required to describe the static (DC) behavior. At high frequencies, similarily six frequency independent elements may be required

\begin{widetext}
\begin{equation}\label{eq:detepsDC}
\det\{ \varepsilon(\omega=0)\}=\varepsilon_{\stext{DC},xx}\varepsilon_{\stext{DC},yy}\varepsilon_{\stext{DC},zz}+2\varepsilon_{\stext{DC},xy}\varepsilon_{\stext{DC},yz}\varepsilon_{\stext{DC},xz}-(\varepsilon_{\stext{DC},xx}\varepsilon_{\stext{DC},yz}^2+\varepsilon_{\stext{DC},yy}\varepsilon_{\stext{DC},xz}^2+\varepsilon_{\stext{DC},zz}\varepsilon_{\stext{DC},xy}^2),
\end{equation}
\begin{equation}\label{eq:detepsinf}
\det\{ \varepsilon_{\infty}\}=\varepsilon_{\infty,xx}\varepsilon_{\infty,yy}\varepsilon_{\infty,zz}+2\varepsilon_{\infty,xy}\varepsilon_{\infty,yz}\varepsilon_{\infty,xz}-(\varepsilon_{\infty,xx}\varepsilon_{\infty,yz}^2+\varepsilon_{\infty,yy}\varepsilon_{\infty,xz}^2+\varepsilon_{\infty,zz}\varepsilon_{\infty,xy}^2).
\end{equation}
\end{widetext}

\noindent According to Eq.~(\ref{eq:sumP}) each element of $\varepsilon$ possesses up to ($N$+1) terms

\begin{equation}\label{eq:epsij}
(\varepsilon)_{ij}=\det\{ \varepsilon_{\infty}\}\hat{e}_{i,\infty}\hat{e}_{j,\infty}+\sum^{N}_{l=1}\varrho_{(l)}\hat{e}_{i,l}\hat{e}_{j,l},\; i,j \in \{x,y,z\}.
\end{equation}

\noindent Hence, $\varepsilon$ is symmetric, invariant under time and space inversion, and a function of frequency $\omega$. The dielectric function tensor in Eq.~(\ref{eq:eps}) has six independent complex-valued parameters, which can be obtained by experiment, for example using generalized spectroscopic ellipsometry~\cite{SchubertPRB61_2000,SchubertIRSEBook_2004,SchubertPRB2016}. Two characteristic optical modes, transverse optical (TO; $\omega_{\stext{TO}}$) and longitudinal optical (LO; $\omega_{\stext{LO}}$), can be obtained, respectively, from the roots of the determinants of $\varepsilon^{-1}$, and $\varepsilon$

\begin{equation}\label{eq:TO&LO}
0=\det\{ \varepsilon^{-1}(\omega_{\stext{TO}})\}, \; 0=\det\{\varepsilon(\omega_{\stext{LO}})\},
\end{equation}

\noindent and a proof for this statement is obtained below. For $0 < \omega < \infty$, one can express the determinant of $\varepsilon$ through a complex-valued function $f$, or $f^{\dagger}$

\begin{equation}\label{eq:f}
\det\{ \varepsilon(\omega)\} = \det\{\varepsilon(\infty)\}+f(\omega)=\det\{\varepsilon(\infty)\}\left(1+f^{\dagger}(\omega)\right),
\end{equation}

\noindent where $f^{\dagger}$ is obtained from $f$ by normalization with $\det\{\varepsilon(\infty)\}$. The sum $1+f^{\dagger}$ contains up to $6(1+$N$)^{3}$ terms. Each term has the following structure

\begin{equation}\label{eq:terms}
\hat{e}_{i,l}\hat{e}_{j,l}\varepsilon_{(l)}\hat{e}_{i,m}\hat{e}_{j,m}\varepsilon_{(m)}\hat{e}_{i,n}\hat{e}_{j,n}\varepsilon_{(n)},
\end{equation}

\noindent where the mode indices are $\{l,m,n\} \in \{``\infty", 1, \dots, N \}$, and the coordinate indices are $\{ i,j\} \in \{x,y,z\}$. In the calculation of the determinant of $\varepsilon$ all terms occur in cyclic permutations of the indices and by alternating plus and minus signatures of all product terms. It is crucial to recognize that in this summation all terms with at least two equal mode indices in $l,m,n$ cancel out. As a result, none of the terms in Eq.~(\ref{eq:epsij}) occur in the sum $1+f^{\dagger}$ with a multiplicity higher than one. This is consequential when the sum $1+f^{\dagger}$ is then factorized into a fraction decomposition with common denominator. The denominator then contains the product over all poles at $(\omega^2_{\stext{TO},(l)}-\omega^2)$ with $ l =1, \dots, N$. This result is obtained straightforward by carrying out all multiplications and by summing all terms in $1+f^{\dagger}$ for arbitrary but fixed $N$. The numerator then presents itself with a polynomial in $\omega^2$ with order equal to $2N$. Hence, the numerator can be factorized according to the Gau\ss-d'Alembert theorem of algebra by which a polynomial $p$ of degree $n$ possesses $n$ roots in the complex plane~\cite{Bronstein}. Hence, for the sum $1+f^{\dagger}$ one expects $N$ roots in $\omega^2$, which are identical then with the squares of the frequencies $\omega_{\stext{LO},(l)}$ in Eq.~(\ref{eq:TO&LO}).  Ignoring broadening here and setting the broadening parameters in Eq.~(\ref{eq:rho}) to zero, the determinant of $\varepsilon$ can be expressed as follows

\begin{equation}\label{eq:LSTfreq}
\det\{ \varepsilon(\omega)\} = \det\{\varepsilon(\infty)\}\prod^{N}_{l=1}\left(\frac{\omega^2_{\stext{LO},(m)}-\omega^2}{\omega^2_{\stext{TO},(l)}-\omega^2}\right).
\end{equation}  

\noindent Obviously, frequencies $\omega_{\stext{TO}}$ as defined in Eq.~(\ref{eq:TO&LO}) are identical with frequencies $\omega_{\stext{TO},(l)}$. 

The factorized form of Function $\det\{ \varepsilon(\omega)\}$ in Eq.~(\ref{eq:LSTfreq}) can be seen as a generalized dielectric response characteristic for any given material regardless of its crystal system. The particular usefulness of this function originates from its zeros, which reveal all LO frequencies of a material under investigation within the spectral range over which the individual components of $\varepsilon(\omega)$ may have been determined, either from computational theory or from experiment. This function can be further factorized into two identities for the monoclinic crystal system, or a set of two different identities for trigonal, tetragonal, and hexagonal crystal systems, or into three identities for the orthorhombic crystal system. The latter three are all identical then for the cubic crystal system. 

Setting $\omega = 0$ in Eq.~(\ref{eq:LSTfreq}), a generalization of the LST relation is then obtained

\begin{equation}\label{eq:LSTgen}
\frac{\det\{ \varepsilon(0)\} }{\det\{\varepsilon(\infty)\}}= \prod^{N}_{l=1}\left(\frac{\omega_{\stext{LO},(l)}}{\omega_{\stext{TO},(l)}}\right)^2,
\end{equation}  

\noindent where the product expands over all $N$ vibration modes contained within Eq.~(\ref{eq:sumP}). Equations~(\ref{eq:LSTfreq}) and~(\ref{eq:LSTgen}) are the central results of this paper. The relations are valid for materials with all crystal systems, and specifically for triclinic. For monoclinic, when without loss of generality the $(x,y)$ plane may be considered as the monoclinic plane, the LST relation factorizes into two identities~\cite{SchubertPRB2016}

\begin{equation}\label{eq:LSTxy}
\frac{\varepsilon_{\stext{DC},xx}\varepsilon_{\stext{DC},yy}-\varepsilon^2_{\stext{DC},xy}}{\varepsilon_{\infty,xx}\varepsilon_{\infty,yy}-\varepsilon^2_{\infty,xy}}=\prod^{N}_{l=1}\left(\frac{\omega_{\stext{LO},l}}{\omega_{\stext{TO},l}}\right)^2,
\end{equation} 

\begin{equation}\label{eq:LSTz}
\frac{\varepsilon_{\stext{DC},zz}}{\varepsilon_{\infty,zz}}=\prod^{K}_{l=1}\left(\frac{\omega_{\stext{LO},l}}{\omega_{\stext{TO},l}}\right)^2,
\end{equation}   

\noindent where $N$ denotes the number of vibration modes whose eigen displacement vectors are all aligned within the monoclinic plane, and $K$ denotes the number of all vibration modes purely polarized along direction $z$. For orthorhombic, the generalized LST relation factorizes into three identities

\begin{equation}\label{eq:LSTxyz}
\frac{\varepsilon_{\stext{DC},x,y,z}}{\varepsilon_{\infty,x,y,z}}=\prod^{K_{x,y,z}}_{l=1}\left(\frac{\omega_{\stext{LO},l}}{\omega_{\stext{TO},l}}\right)^2,
\end{equation}   

\noindent where $K_{x,y,z}$ denotes the number of vibration modes whose eigen displacement vectors are aligned parallel to the major orthorhombic lattice axes, which may be aligned with $x,y,z$, respectively. The static and high frequency dielectric constants, $\varepsilon_{\stext{DC},x,y,z}$ and $\varepsilon_{\infty,x,y,z}$, relate to the three major orthorhombic axes, respectively. For trigonal, tetragonal, and hexagonal, $K_{x,y} = K_{\perp}$ and $K_{z}=K_{||}$ denote the numbers of vibration modes perpendicular and parallel to the lattice $\mathbf{c}$ axis, respectively, when $\mathbf{c}$ is aligned parallel to $z$, and the static and high frequency dielectric constants are $\varepsilon_{\stext{DC},x,y}=\varepsilon_{\stext{DC},\perp},\; \varepsilon_{\stext{DC},z}=\varepsilon_{\stext{DC},||}$ and $\varepsilon_{\infty,x,y}=\varepsilon_{\infty,\perp},\; \varepsilon_{\infty,z}=\varepsilon_{\infty,||}$. For cubic materials, taking the third root of Eq.~(\ref{eq:LSTgen}) because of triple degeneracy, the original LST relation is recovered in Eq.~(\ref{eq:LSTproduct}), identical to Eq.~(\ref{eq:LSTsimple}) for $K=1$ 

\begin{equation}\label{eq:LSTis}
\frac{\varepsilon_{\stext{DC}}}{\varepsilon_{\infty}}=\prod^{K}_{l=1}\left(\frac{\omega_{\stext{LO},l}}{\omega_{\stext{TO},l}}\right)^2,
\end{equation}   

\noindent where $K$ denotes the number of vibration modes within the cubic material. 

Finally, a generalized oscillator strength which combines the polarizability of all long wavelength active vibration modes in a given sample can be derived from Eq.~(\ref{eq:LSTgen})~\cite{Klingshirn95,Dressel_2002,GrundmannBook}

\begin{equation}\label{eq:genf}
f=\sqrt[3]{ \left(\frac{\det\{ \varepsilon(0)\} }{\det\{\varepsilon(\infty)\}}-1\right)\prod^{N}_{l=1}\omega^2_{\stext{TO},(l)}},
\end{equation}  

\noindent where the product runs over all $N$ vibration modes. The appearance of the third root in Eq.~(\ref{eq:genf}) reflects the fact that the derivation comprises all modes in all three dimensions. For example, a material with cubic crystal system with $K$ distinct vibration frequencies is accounted for in Eq.~(\ref{eq:epsij}) by adding three equal functions $\varrho_{(l)}$ for every vibration mode, where the three functions $\varrho_{(l)}$ are assigned with a set of orthogonal eigenvectors, for example, aligned along $x,y,z$. The product then extends to $N=3K$ and terms in $\omega_{\stext{TO},(l)}$ occur three times. The value of Eq.~(\ref{eq:genf}) consists in the possibility to express a generalized oscillator strengths in units of the vacuum permittivity, $\varepsilon_0$, which can be calculated without explicit knowledge of the LO frequencies.

In a recent experiment, the dielectric function tensor components of single crystal monoclinic $\beta$-Ga$_2$O$_3$ were measured by generalized ellipsometry in the long wavelength spectral range~\cite{SchubertPRB2016}. All long wavelength active phonon modes predicted by theory were detected as well as their eigen vectors within the monoclinic plane. The tensors of the static and high frequency dielectric constants were determined from experiment and the generalized form of the LST relation was found fulfilled accurately, lending experimental support to the findings reported here. No other experimental data appear to be available for materials with monoclinic or triclinic crystal systems, and future experiments may provide further tests of the LST relations provided here. 

A generalization of the Lyddane-Sachs-Teller relation is derived for polar vibrations in materials with monoclinic and triclinic crystal systems. The generalization is derived from an eigen displacement vector summation approach, which is equivalent to the microscopic Born-Huang description of polar lattice vibrations. The generalized relation is found valid for monoclinic $\beta$-Ga$_2$O$_3$, where accurate experimental data became available recently from a comprehensive generalized ellipsometry investigation. Data for materials with triclinic crystal systems can be measured by generalized ellipsometry as well, and are anticipated to become available soon and results can be compared with the generalized relation discussed here.

\section{Acknowledgments} This work was supported in part by the National Science Foundation (NSF) through the Center for Nanohybrid Functional Materials (EPS 1004094), the Nebraska Materials Research Science and Engineering Center (DMR 1420645), and award CMMI 1337856.

\bibliography{CompleteLibrary}

\end{document}